\documentclass{ws-ijmpa}

\begin{document}

\markboth{Dimitri Bourilkov}
{THE CAVES/CODESH Projects}

%
\catchline{}{}{}{}{}
%

\title{THE CAVES PROJECT\\
Collaborative Analysis Versioning Environment System\\
THE CODESH PROJECT\\
COllaborative DEvelopment SHell\\
\vspace{0.2cm}
\raggedleft \footnotesize{GriPhyN 2004-73; physics/0410226, October 2004}
}

\author{\footnotesize Dimitri Bourilkov}

\address{Physics Department, University of Florida, P.O. Box 118440\\
Gainesville, FL 32611, USA
}

\maketitle

\pub{Received 22 October 2004}{}

\begin{abstract}
A key feature of collaboration in science and software development is
to have a {\em log} of what and how is being done - for private use and
reuse and for sharing selected parts with collaborators, which most often
today are distributed geographically on an ever larger scale.
Even better if this log is {\em automatic}, created on the fly while
a scientist or software developer is working in a habitual way, without
the need for extra efforts. The {\tt CAVES} and {\tt CODESH} projects
address this
problem in a novel way, building on the concepts of {\em virtual state}
and {\em virtual transition} to provide an automatic persistent logbook
for sessions of data analysis or software development in a collaborating group.
A repository of sessions can be configured dynamically to record and make
available the knowledge accumulated in the course of a scientific or software
endeavor. Access can be controlled to define logbooks of private sessions and
sessions shared within or between collaborating groups.

\end{abstract}

\section{Introduction}

Have you sifted through paper folders or directories on your computer,
trying to find out how you produced a result a couple of months (or years)
ago? Or having to answer a question while traveling, with your folders
safely stored in your office? Or a desperate collaborator trying to
reach you about a project detail while you are hiking in the mountains?
It happened many times to me, and there must be a better way. 

\section{Automatic Logbooks}

In this paper the possibility to create an automated system for recording
and making available to collaborators and peers the knowledge accumulated
in the course of a project is explored. The work in a project is atomized
as {\em sessions}. A session is defined as a transition $T$ from an initial 
$|I>$ to a final $|F>$ state:
$$|F> = T |I>.$$
In this notation the work done during a session is represented by the
transition operator $T$. The length of a session is limited by the decision
of the user to record a chunk of activity.

A state can be recorded with different level of detail determined by the
users:
$$|State> = |Logbook part, Environment>.$$
The system records automatically, in a persistent way, the logbook part for
future use. Information about the environment can be logged as well, but
it is not stored in its entirety (e.g. the operating system).
The environment is considered as available
or provided by the collaborating group. The splitting between logbook
and environment parts is to some extent arbitrary. We will call it a
{\em collaborating contract}, in the sense that it defines the responsibilities
of the parties involved in a project.\footnote{In these terms a computing grid
can be viewed as a collection of sites with different contracts. An application
selects the sites with contracts suitable for running there.}
A good splitting is characterized by the transition operator $T$ acting in a
 meaningful way on the logged part without affecting substantially the
environment.

If the level of detail is sufficient, each final state $|F>$ can be reproduced
at will from the log of the initial state $|I>$ and the transition $T$. A final
state for one transition can serve as initial state for a new transition, 
enabling us to build arbitrarily long chains. The ability to reproduce
states bring us to the notion of {\em virtual state} and
{\em virtual transition}, in the sense that we can record states that existed
in the past and can recreate them on demand in the future. Virtual states
serve as {\em virtual checkpoints}, or delimiters for {\em virtual sessions}.

In a traditional programming language, the user typically codes the
transition $T$, i.e. provides in advance the source of the program for
producing a given final state, following the syntax for the language used.
The key idea in our approach is that the user works in a habitual way
and the log for the session is created {\em automatically, on the fly,} while
the session is progressing. There is no need per se to provide any code in
advance, but the user can execute preexisting programs if desired. When a
piece of work is worth recording, the user logs it in the persistent session
repository with a unique identifier.
\footnote{The virtual data idea explores similar concepts, putting more
emphasis on the transition - abstract transformation and concrete
derivations, coded by the users in a virtual data language.~\cite{chimera}}

Let us illustrate this general framework with an example. We want to log
the activities of users doing any kind of work on the command line e.g.
in a UNIX shell. We start a session by recording details about the initial
state e.g. the key-value pairs for all defined environment variables and
aliases. Then we give commands to the shell 
(e.g. ls, cd, cp, mv, find, grep etc). During the session we could run some
existing user shell scripts (e.g. in csh, bash, python, perl ...),
possibly providing input/output parameters when invoking them.
If we collect and store all commands and the source code of all executed
scripts as used during the session, we have automatically produced a log
for the transition $T$. Optionally we can record the environment variables and
aliases at the end of the session. Later we may change any of the used scripts,
even if they still have the same filenames, delete or accidentally lose some
of them, forget which parameters were used or why. When we or a collaborator
want to reproduce the session, we create a new sandbox (clean slate) and
download from the repository the log for a given unique identifier. So we have
both the commands and the scripts, as well as the input/output parameters,
``frozen'' as they were at the time of the session, and can reproduce
the results.

The automatic logbooks can be augmented by annotations. If the users
consider it helpful, they can provide additional information, just a
couple of words or a whole paragraph, to describe the session.
The annotations can be browsed by the members of a group, thus enhancing
the collaborative experience both for experts and newcomers, and making the
selection of sessions of interest much easier. When desirable,
information from different phases of a project can easily be shared 
with other groups or peers.

We stressed already the value of complete logs. In the heat of an
active session, when there is no time or need to be pedantical,
users may see merit in storing sometimes also partial logs, a classical
example being a program with hidden dependencies, e.g. the calling of
a program or reading of a file within a program, not exposed externally.
In this case, the final state will not be reproducible, but at least the
log will point what is missing. Or the users may even store a non-functional
sequence of actions in the debugging phase for additional work later,
even without producing a well defined final state. Our system
should be able to support partial logging, leaving the spectrum of
possible use cases to the imagination of the collaborators.

\section{The CODESH/CAVES Projects - Working Implementations
of Automatic Logbooks}

The {\tt CODESH} and {\tt CAVES}~\cite{Bourilkov:2004nj} projects take a
pragmatic approach in assessing the
needs of a community of scientists or software developers by building series
of working prototypes with increasing sophistication. By extending with
automatic logbook capabilities the functionality of a typical {\tt UNIX}
shell (like tcsh or bash) - the {\tt CODESH} project, or a popular analysis
package as {\tt ROOT}~\cite{root} - the {\tt CAVES} project, these
prototypes provide an easy and habitual entry point for researchers to
explore new concepts in real life applications and to give
valuable feedback for refining the system design. Our goal is to stay close
to the end users and listen carefully to their needs at all stages of a
developing project. We have found that
proceeding in this way helps to approach the optimal architecture.

Both projects use a three-tier architecture, with the users sitting at
the top tier and running what looks very much like a normal shell or
identical to a {\tt ROOT} session, and having extended capabilities, which are
provided by the middle layer. This layer is coded in {\tt Python} for
{\tt CODESH} or in {\tt C++} for {\tt CAVES}, by inheriting from the
class which handles the user input on the command line. The implemented 
capabilities are similar to the example in the previous section.
There is no need to learn yet another programming language, and our goal is
simplicity of design, keeping the number of commands and their parameters to
the bare minimum needed for rich and useful
functionality.~\cite{Bourilkov:2004nj}
The lower tier provides the persistent back-end. The first implementations
use a well established source code management system - the
Concurrent Versions System {\tt CVS}.
It is well suited to provide version control for a rapid development by
a large team and to store, by the mechanism of tagging releases, many versions
so that they can be extracted in exactly the same form even if modified,
added or deleted since that time. The {\tt CVS} tags assume the role of
unique identifiers for virtual sessions.
More back-ends based on Web, Grid and other services are under
development.~\cite{gae}

\section{Outlook}

In this paper we have outlined the main ideas driving the
{\tt CODESH} and {\tt CAVES}
projects for exploring automatic logbook concepts in scientific collaboration
and software development.
The decomposition of e.g. a typical analysis or shell session shows that
the automatic logbook / virtual session approach bears great promise for
qualitatively enhancing the collaborative work of research and software groups
and the accumulation and sharing of knowledge in todays complex, large
scale scientific / software environments. The confidence in
results and their discovery and reuse grows with the ability to automatically
log and reproduce them on demand.

We have built first functional systems providing automatic
logging in a typical working session. The systems have been demonstrated
successfully at Supercomputing 2003, the {\tt ROOT} Workshop and
the DPF Conference in 2004. Public releases are available
for interested users, which are encouraged to contact the
author.\footnote{bourilkov@phys.ufl.edu}

\section*{Acknowledgments}

I would like to thank all members from the Florida group of Paul Avery
for the creative atmosphere which helped bring this work to fruition.
The study is supported in part by the United States National Science Foundation
under grants NSF ITR-0086044 (GriPhyN) and NSF PHY-0122557 (iVDGL).

\end{document}